# Physics-constrained 3D Convolutional Neural Networks for Electrodynamics

Alexander Scheinker* and Reeju Pokharel
Los Alamos National Laboratory
*Electronic Email: ascheink@lanl.gov

**Abstract**
We present a physics-constrained neural network (PCNN) approach to solving Maxwell's equations for the electromagnetic fields of intense relativistic charged particle beams. We create a 3D convolutional PCNN to map time-varying current and charge densities **J**(**r**,t) and ρ(**r**,t) to vector and scalar potentials **A**(**r**,t) and φ(**r**,t) from which we generate electromagnetic fields according to Maxwell's equations: **B** = ∇×**A**, **E** = −∇φ −∂**A**/∂t. Our PCNNs satisfy hard constraints, such as ∇ · **B** = 0, by construction. Soft constraints push **A** and φ towards satisfying the Lorenz gauge.

## INTRODUCTION

Electrodynamics is ubiquitous in describing physical processes governed by charged particle dynamics including everything from models of universe expansion, galactic disks forming cosmic ray halos, accelerator-based high energy X-ray light sources, achromatic metasurfaces, metasurfaces for dynamic holography and on-chip diffractive neural networks, down to the radiative damping of individual accelerated electrons [1-21].

Despite widely available high-performance computing, numerically calculating relativistic charged particle dynamics is still a challenge and an open area of research for large collections of particles undergoing collective effects in dynamics involving plasma turbulence [22], space charge forces [23,24], and coherent synchrotron radiation [25,26]. For example, the photo-injectors of modern X-ray free electron lasers such as the LCLS, SwissFEL, and EuXFEL and plasma wakefield accelerators such as FACET-II can produce high quality intense bunches with few picosecond rms lengths of up to 2 nC charge per bunch that are accelerated and squeezed down to lengths of tens to hundreds of femtoseconds [27-32]. At low energy near the injector, the 6D phase space ($x,y,z,p_x,p_y,p_z$) dynamics of such bunches are strongly coupled through collective space charge (SC) forces. At higher energies, especially in bunch compressors where charged particle trajectories are curved through magnetic chicanes the dynamics are coupled through collective coherent synchrotron radiation (CSR).

A 2 nC bunch contains $N \approx 1.25 \times 10^{10}$ electrons for which calculating exact individual particle to particle SC and CSR interactions is a computationally expensive $O(N^2)$ process. For SC calculations, an $O(N^2)$ process, such as the *SpaceCharge3D* routine in the particle dynamics simulation code General Particle Tracer (GPT) may be necessary for intense low energy beams near the injector where the longitudinal (*z*) velocities of individual particles in a bunch have a large variation and are comparable to transverse (*x,y*) velocities [33,34]. For relativistic particles, many conventional approaches for SC calculations greatly reduce the number of required calculations by utilizing particle-in-cell methods with macro-particles, such as the *SpaceCharge3DMesh* routine in GPT. For CSR, relativistic state-of-the-art 3D CSR calculations still rely on a full set of point-to-point calculations [35].

A charged particle's electromagnetic Lagrangian is

$$L = -\frac{mc^2}{\gamma} + e\mathbf{v} \cdot A - e\varphi, \qquad \gamma = \frac{1}{\sqrt{1-v^2/c^2}}, \qquad (1)$$

where *e* is the particle's charge, c is the speed of light, $v = |\mathbf{v}|$, and A and φ are the vector and scalar potentials, respectively, which define the magnetic (**B**) and electric (**E**) fields as

$$\mathbf{B} = \nabla \times \mathbf{A}, \qquad \mathbf{E} = -\nabla\varphi - \frac{\partial \mathbf{A}}{\partial t}, \qquad (2)$$

For which the relativistic Lorentz force law is

$$\frac{d\mathbf{p}}{dt} = e(\mathbf{E} + \mathbf{v} \times \mathbf{B}), \qquad \mathbf{p} = \gamma m \mathbf{v}. \qquad (3)$$

The **E** and **B** dynamics are coupled and depend on current and charge densities **J** and ρ as described by Maxwell's equations

$$\nabla \cdot \mathbf{E} = \frac{\rho}{\varepsilon_0}, \qquad \nabla \cdot \mathbf{B} = 0, \qquad (4)$$

$$\nabla \times \mathbf{E} = -\frac{\partial \mathbf{B}}{\partial t}, \qquad \nabla \times \mathbf{B} = \mu_0 \left( \mathbf{J} + \varepsilon_0 \frac{\partial \mathbf{E}}{\partial t} \right). \qquad (5)$$

A typical approach to numerically solving Eq. 1-5 starts with initial charge ρ(*x,y,z,t* = 0) and current profiles **J**(*x, y, z, t* = 0) and their rates of change as well as any external electric and magnetic fields **E**$_{ext}$(*x,y,z,t* = 0), **B**$_{ext}$(*x,y,z,t* = 0) and their rates of change, which may be produced by mag- nets and radio-frequency resonant acceleration cavities as is typical in high intensity charged particle accelerators. The total electromagnetic fields are then calculated as the sum of the external fields and the self-fields produced by the current and charge densities themselves according to Eq. 4, 5. The initial fields apply a force on the particles causing a change in momentum and position, as defined by Eq. 3. The most computationally expensive part of the process is the calculation of the self-fields generated by the particle distribution.

In this work, we introduce a physics-constrained neural net- work (PCNN) approach to solving Maxwell's equations for the self-fields generated by relativistic charged particle beams. For example, for the problem of mapping current density J to an estimate $\hat{\mathbf{B}}$ of the associated magnetic field B we build Eq. 2 into the structure of our NN and generate the vector potential $\hat{\mathbf{A}}$, which defines the magnetic field as

$$\hat{\mathbf{B}} = \nabla \times \hat{\mathbf{A}} \implies \nabla \cdot \hat{\mathbf{B}} = \nabla \cdot (\nabla \times \hat{\mathbf{A}}) = 0, \qquad (6)$$

which satisfies the physics constraint by construction.

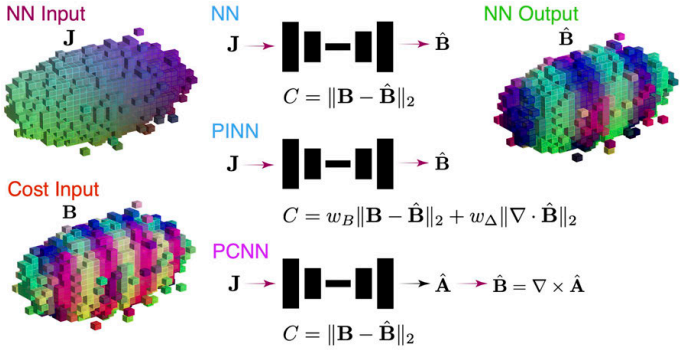

Fig 1: Various NN approaches to generate a magnetic field **B^** (an estimate of **B**) from current density **J** are shown.

Neural networks (NN) are powerful machine learning (ML) tools which can extract complex physical relationships directly from data and have been used for speeding up the studies of complex physical systems [36-43]. Incredibly powerful and flexible physics-informed neural networks (PINNs), which include soft constraints in the NN's cost function, have been developed and have shown great capabilities for complex fluid dynamics simulations [44], material science [45], for symplectic single particle tracking [46], for learning molecular force fields [47], and for large classes of partial differential equations [48-50].

For the problem of mapping current density J to an estimate B^, the PINN approach is to train a neural network with a cost function defined as

$$C = w_B \iiint |B - \hat{B}|^2 dV + w_\nabla \iiint |\nabla \cdot \hat{B}|^2 dV$$
$$= w_B \|B - \hat{B}\|_2 + w_\nabla \|\nabla \cdot \hat{B}\|_2, \quad (7)$$

where the first term depends on magnetic field prediction accuracy and the second term penalizes violation of the physics constraint $\nabla \cdot \hat{B} = 0$, as shown in Figure 1. However, with soft PINN-type constraints there is no guarantee that the constraints are always satisfied, which is in contrast to the hard constraints implemented in our approach, which guarantee that constraints are not violated within numerical and finite discretization limits. Furthermore, when utilizing PINN-type soft constraints there is a tradeoff between the minimization of the two terms in Eq. 7 based on the choice of weights $w_B$ and $w_\Delta$. Intuitively this tradeoff can be understood by the fact that the easiest way for a neural network to satisfy $\nabla \cdot \hat{B} = 0$ is $\hat{B} \equiv C$ for any constant $C$. For hard constraints there is no such tradeoff, the cost function only penalizes field accuracy and the constraint itself is built into how the field is constructed. In our PCNN approach, our cost function is simply

$$C = \|B - \hat{B}\|_2, \quad (8)$$

and there is no tradeoff between reconstruction accuracy and physics constraint enforcement.

This is important because when simulating charged particle dynamics, great care must be taken to satisfy the physics constraints as defined by Eq. 1-5. It is very important to enforce well known beam properties such as phase space volume-preserving symplectic maps that satisfy Liouville's theorem so that the beam dynamics are self-consistent [51-56]. Results on physics informed NNs with hard constraints have mostly focused on fluid dynamics and climate modeling and are much more limited than PINN approaches [57-61].

## PHYSICS-CONSTRAINED NEURAL NETWORKS

In Figure 1 we summarize three NN approaches: 1) a NN approach without physics constraints, 2) a PINN approach with soft constraints, and 3) our PCNN approach. We demonstrate our PCNN method with numerical studies of relativistic (5 MeV), short ($\sigma_t$ = 800 fs), high charge (2 nC) electron bunches represented by $N$ = 50 million macro particles. We utilize the charged particle dynamics simulation code General Parcticle Tracer (GPT) with 3D space charge forces [33,34]. The charged particle distributions were simulated for 1.2 ns with all of the data saved to a file at each $\Delta t$ = 12 ps interval so that the beam was displaced 0.36 m over 100 saved steps.

The Figure 2 (A) shows the $x$ and $y$ trajectories of 10000 random particles sampled from the bunch distribution over the entire 100 saved steps as the beam is compressed by a 0.5 T solenoid magnet whose $B_z$ field is shown in green. Only the first 75 steps, shown in black, were used for training and the final 25 steps were used for testing, shown in red. Figure 2 (B) shows the (x,y) and (x,z) projections of the electron bunch density at steps 0 and 74.

The training beam we have created is designed to have multiple length scales in order to help the trained PCNN generalize to new unseen distributions. We have created several closely spaced Gaussian bunches of varying $\sigma_x$ and $\sigma_y$ as seen in the (x,y) projection of step 0. Furthermore as seen in the (x,z) projection the beam has an overall bunch length of $\sigma_z$ = 800 μm with density fluctuations of various $\sigma_z$ along the length of the beam. By step 75 the beam has been overcompressed in the (x,y) plane as seen by the (x,y) projection and the beam density has started to spread in the z direction due to space charge forces.

At each time step we generate discrete versions of J, ρ, E, and B by breaking up the 2.4 mm × 2.4 mm × 4.4 mm volume which is co-moving with the center of the beam into a 128×128×128 pixel cube with sides of length $\Delta_x$ = 18.9 μm, $\Delta_y$ = 18.9 μm, $\Delta_z$ = 34.6 μm, and averaging over all of the macroparticles in each cube. We compare the three neural network approaches to map J to B, as shown in Figure 1: 1) A standard NN using (8) as the cost function for training, 2) A PINN using (7) as the cost function, and 3) A PCNN us- ing (8) as the cost function with the physics constraint built into the structure of the ML approach. The NN, PINN, and PCNN are able to achieve similar errors on the training data as they all use a similar 3D convolutional neural network (CNN) encoder-decoder architecture, as shown in Figure 3.

There is however an important distinction in terms of neural network size when comparing the NN, PINN, and PCNN approaches. The PCNN is actually smaller while achieving better test data results and a much smaller violation of the physics constraint. For the NN and PINN all three components ($J_x$, $J_y$, $J_z$) of **J** must be used as inputs to generate all three components ($\hat{B}_x$, $\hat{B}_y$, $\hat{B}_z$) of **B̂**. Therefore, the inputs and outputs of this 3D CNN are objects of size 128×128×128×3 and the input-output mapping of this 3D CNN, $N_B$, is given by

$$\{J_x, J_y, J_z\} \to N_B \to \{\hat{B}_x, \hat{B}_y, \hat{B}_z\}. \quad (9)$$

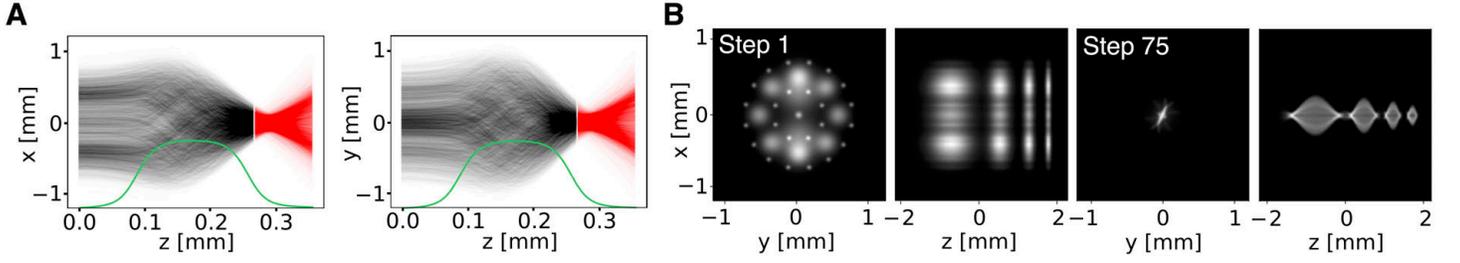

Fig 2: (A) 10000 randomly chosen macroparticle trajectories as they are compressed by a solenoid magnet (green). Initial 75 steps are used for training (black) and final 25 for testing (red). (B) Initial and compressed beam charge density (x,y) and (x,z) projections.

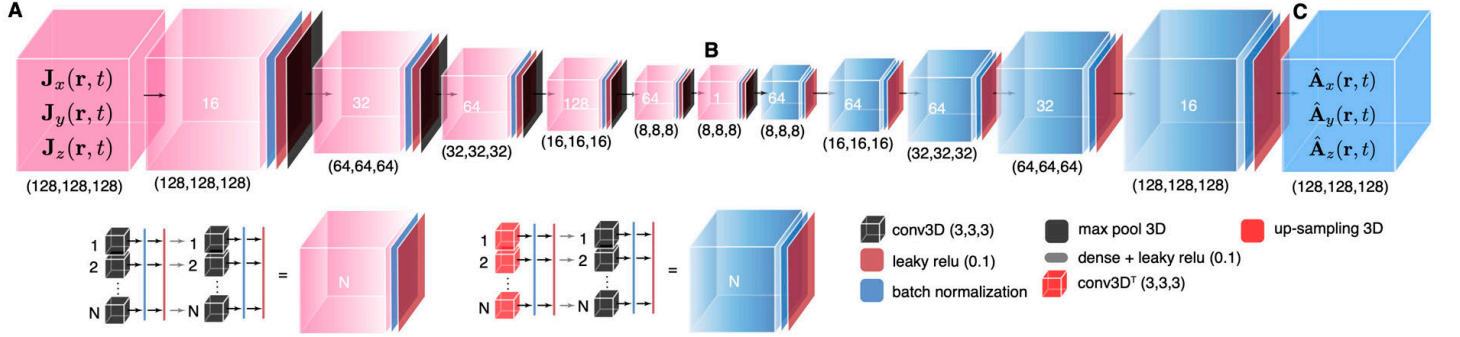

Fig 3: (A) A deep convolutional neural network-based encoder-decoder architecture is used with a 128 × 128 × 128 pixel 3D input. (B) The relatively small 8 × 8 × 8 latent space at the center of the network ensures that each pixel is a function of every other pixel in the 3D input. (C) The latent space volume is then expanded back up to the original size in the generative half of the network.

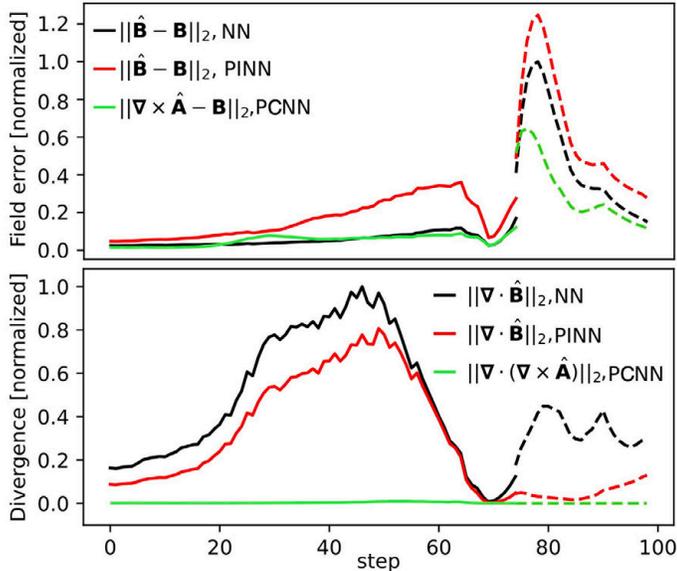

Fig 4: Comparison of mean absolute errors of field errors and of field divergence for the three approaches, normalized by the maximum error obtained by the standard NN approach. For the PINN approach there is a tradeoff between the accuracy of the prediction and violation of the soft constraint.

However, for the PCNN we are generating an estimate of **A**, which satisfies

$$A(r,t) = \frac{\mu_0}{4\pi} \iiint \frac{J(r',t)}{|r-r'|} d^3r', \quad (10)$$

and therefore, each component of A only depends on the corresponding component of J and the dependence of each component of **A** on **J** has the same functional form. Therefore, we are able to train just a single one-channel 3D CNN, $N_A$, which takes only one input at a time of size 128×128×128×1, and the input-output mapping is

$$J_\blacksquare \to N_A \to \hat{A}_\blacksquare, \quad \blacksquare \in \{x, y, z\}, \quad (11)$$

as shown in Figure 3. A single channel 3D CNN approach results in a smaller network with fewer weights and also effectively triples the amount of training data seen by a single network which helps with generalization. Furthermore, the memory requirement is significantly smaller, allowing for larger batch sizes and bigger overall 3D volumes, which will be especially important for enabling future work utilizing even larger 3D objects of up to $1024^3$ pixels. When utilizing even the most expensive GPU workstations, which can comfortably sit in one's office, going up to volumes of $1024^3$ pixels uses up so much GPU memory that the number of 3D convolutional layers that can be utilized in a 3D CNN is greatly diminished and going beyond this will probably require the use of HPC clusters of many GPUs.

For ∇×A we estimate ∂/∂x at pixel $(i,j,k) \in \{1,...,128\}^3$ as

$$\frac{\partial A_\blacksquare^{ijk}}{\partial x} = \frac{A_\blacksquare^{(i+1,j,k)} - A_\blacksquare^{(i-1,j,k)}}{2\Delta_x} + O(\Delta_x^2), \quad (12)$$

where $\blacksquare \in \{x, y, z\}$, and similarly for ∂/∂y and ∂/∂z. The difference computation in Eq. 12 is implemented as a single non-trainable 3D convolutional layer with custom-designed weight tensor, $W_{\partial x}$, such that the 3D convolution applied to a volume gives the partial derivative $V \to V * W_{\partial x} = \partial V/\partial x$, where each pixel $V^{ijk}$ in the input volume is replaced by a local 3×3×3 convolution $V^{ijk} \to \sum_{i'j'k'} V^{i'j'k'} W_{\partial x}{}^{i'j'k'}$ with $W_{\partial x}$ defined as

$$W_{\partial x} = \begin{bmatrix} 0 & 0 & 0 \\ 0 & -\frac{1}{2\Delta_x} & 0 \\ 0 & 0 & 0 \end{bmatrix} \times \begin{bmatrix} 0 & 0 & 0 \\ 0 & 0 & 0 \\ 0 & 0 & 0 \end{bmatrix} \times \begin{bmatrix} 0 & 0 & 0 \\ 0 & \frac{1}{2\Delta_x} & 0 \\ 0 & 0 & 0 \end{bmatrix}, \quad (13)$$

and similarly for $W_{\partial y}$ and $W_{\partial z}$. With this approach all of our computations are performed within a 3D CNN utilizing automatically differentiable GPU-enabled TensorFlow libraries [62]. Evaluation of a single forward pass

$$J \to N_A \to \widehat{A} \to \widehat{B} = \nabla \times \widehat{A} \quad (14)$$

requires only milliseconds for large 128 × 128 × 128 pixel 3D objects when running our 3D CNN on two NVLinked 80 GB RAM NVIDIA A100 GPUs in a high performance desk-top workstation. By comparison, the *SpaceCharge3DMesh* method of GPT for these simulations running on a high-performance desktop with dual Intel Xeon Platinum 8276 CPUs with 28 cores and 56 threads per CPU required more than one minute per calculation of a $\Delta t$ = 12 ps interval.

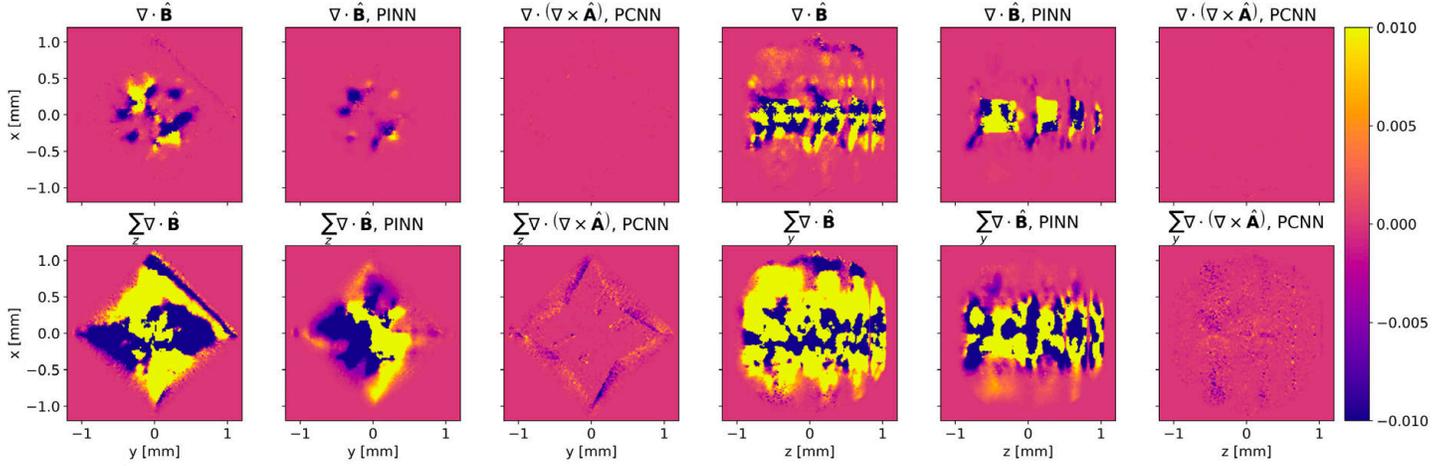

Fig 5: The PINN approach can be made to have smaller divergence by increasing the divergence weight $w_\Delta$ in Eq. 7, but the tradeoff is decreased field prediction accuracy. The PCNN approach has no such tradeoff as the entire cost function is based on the accuracy of the field reconstruction and the constraint is built into the structure of the approach. Here we show several views of $\nabla \cdot B$ for the beam at step 90, far beyond the end of the training set. Top Row: Slices of $\nabla \cdot B$ in the (x,y) plane at z = 0 and in the (x,z) plane at y = 0 are shown for the three methods. Bottom Row: $\nabla \cdot B$ is shown projected onto the (x, y) and (x, z) planes by summing over all z and y values, respectively. The sharp rectangular boundaries of high divergence values are caused by numerical issues at locations where the beam density suddenly drops to zero. This is due to the fact that a slight non-zero background of particles was initially generated within a rectangular volume as the initial conditions for the GPT simulation and those particles have now been compressed and rotated by the solenoid magnet.

## RESULTS

The results of training on the first 75 steps and then testing on the last 25 are shown in Figure 4. While all of the networks learn how to reproduce the magnetic field, the PCNN does the best job of respecting the physics constraint $\nabla \cdot B = 0$. The PINN, depending on the weights chosen in $w_B$, $w_\Delta$ in Eq. 7, can make either one of the costs arbitrarily small, but there is an inherent tradeoff between the two. The PCNN simultaneously reconstructs the fields with high accuracy while always satisfying hard physics constraints. The PCNN also performs better on both reconstruction accuracy and physics constraints on the unseen test data.

Our approach is to generate a general vector field F and then calculate the curl $\nabla \times \mathbf{F}$ and train the 3D CNN such that $\nabla \times \mathbf{F}$ matches the magnetic field **B**. By doing this we have built in the hard constraint that the only representations of magnetic fields that we can construct, **B^**, are the curl of a vector field **F**. For a twice continuously differentiable vector field F this guarantees

$$\nabla \cdot \widehat{\mathbf{B}} = \nabla \cdot (\nabla \times \mathbf{F}) = 0. \quad (15)$$

Once the CNN is trained and **B^** represents a magnetic field, we interpret **F** as an estimate of the vector potential **A^**, due to the fact that **B^** = $\nabla \times$ **A**. Note that due to the finite discretization, which in our case is 128×128×128, and numerical limitations, our vector fields are not perfectly continuous or differentiable and can slightly violate $\nabla \cdot$ **B^** = 0 due to our discretized implementation. This is especially apparent in low-density regions near the edges of the beam where the current and vector densities can discontinuously drop to zero from one pixel to the next, introducing numerical errors in the derivatives. This can be seen most clearly in the bottom row of Figure 6. In Figure 5 the (x, y) and (x, z) projections of slices through the center z=0 and y=0 of the beam are shown for $\nabla \cdot$**B^** for the three approaches. We also project $\nabla \cdot$ **B^** onto the (x, y) plane by summing over all z values and also where we project onto the (x, z) plane by summing over all y values.

After we first generated our particle distributions in GPT and ran the simulations, we had to choose a finite volume to discretize into a 128×128×128 pixel grid to create the 3D density objects for the 3D CNN. We chose dimensions $\Delta x \Delta y \Delta z$ of approximately 1 mm$^3$ because that captured the vast majority of the beam in the (x, y, z) dimensions throughout its evolution through the solenoid. One limitation of this approach which now becomes evident is the cut off the low-density particle regions outside of the chosen volume.

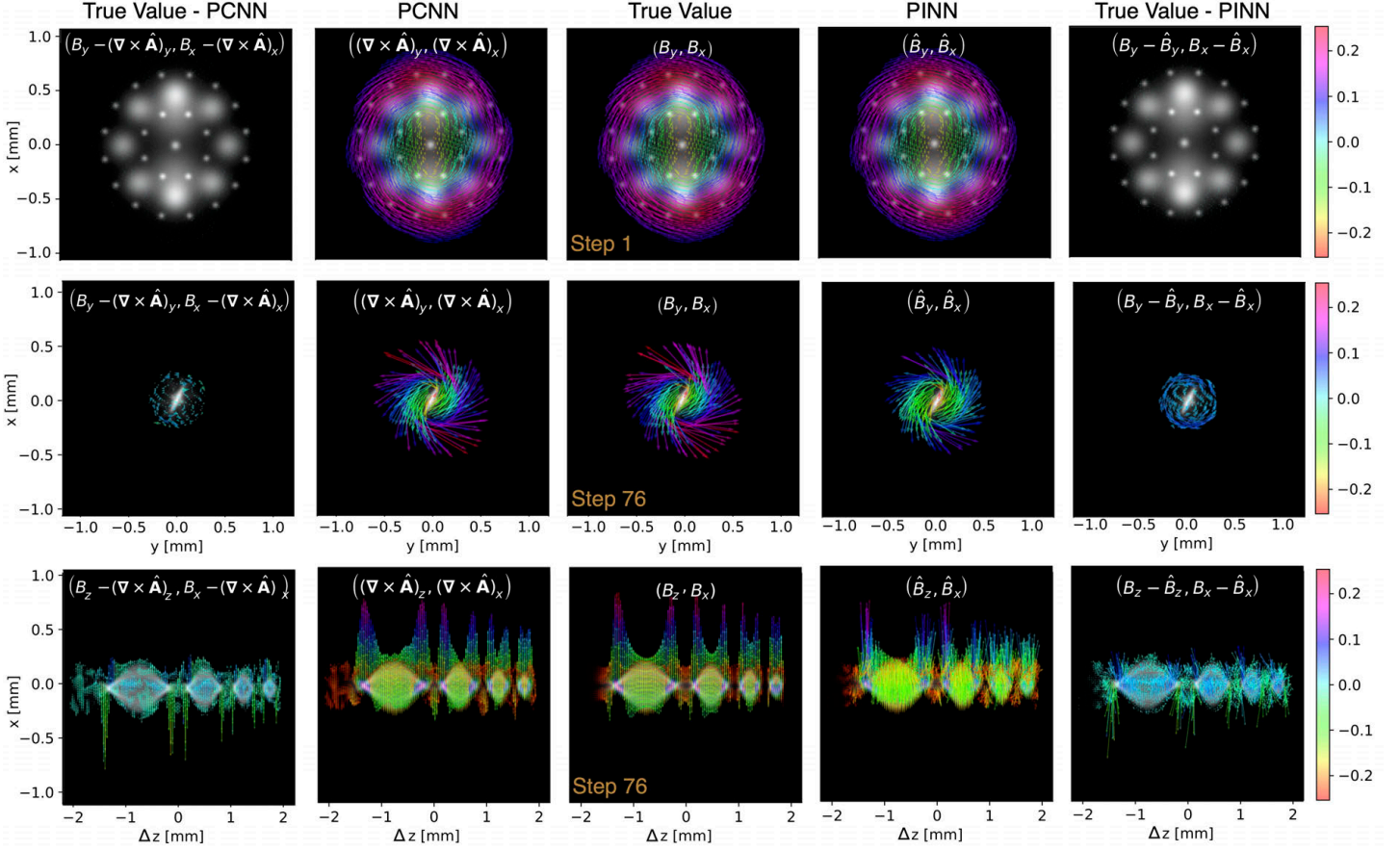

Fig 6: The top row is the initial beam state with the (x,y) projection of $\hat{\mathbf{B}}$ as generated by the PINN and PCNN shown along with **B** and the differences plotted over the (x,y) projection of the beam's charge density ρ. For training data the methods perform equally well in field reconstruction accuracy. In the middle row we see the first step (76) beyond the training data set and an immediate drop in the accuracy of the PINN. In the third row the (x,z) projections of step (76) are shown, the roughness of the (x,z) projection shows intuitively how the PINN matches the B field in a mean squared error sense, but violates the constraint $\nabla \cdot \mathbf{B} = 0$.

Therefore our initial particle distribution can be thought of as an intense beam surrounded by a cube-shaped halo of diminishing density. This is due to the fact that we defined our initial beam in terms of Gaussian distributions without any hard cut-offs. Once this cube shaped region begins to travel through the solenoid it is rotated and squeezed, resulting in regions of non-zero and zero density that have sharp straight contours, as can be seen in the bottom part of Figure 5 and cause numerical problems for calculating derivatives. The most obvious mitigation for this would be to create a mask that cuts off all field calculations related to the beam beyond some minimal cutoff density. Despite this limitation, which each 3D CNN-based approach will suffer, the PCNN approach can be seen to be more accurate than NN without constraints and also than the PINN approach. In Figure 6 we compare PINN and PCNN predictions for two states of the beam, one within the training data set for which both are highly accurate and one beyond the training set where the accuracy quickly drops off. The next step is to add a prediction $\hat{E}$ of the electric field E. We generate φ from ρ via a second neural network $N_\varphi$ which gives the mapping

$$\rho \rightarrow N_\varphi \rightarrow \hat{\varphi}. \qquad (16)$$

As above, we approximate ∂/∂t as

$$\frac{\partial \mathbf{A}}{\partial t} = \frac{A(t+\Delta_t) - A(t-\Delta_t)}{2\Delta_t} + O(\Delta_t^2), \qquad (17)$$

where $\Delta_t = 1.2 \times 10^{-11}$. After $\partial \hat{\mathbf{A}}/\partial t$ is calculated, a single forward pass for $\mathbf{E}^\wedge$ is given by

$$\{\rho, \mathbf{J}\} \rightarrow \{N_\varphi, N_A\} \rightarrow \{\hat{\varphi}, \hat{A}\} \rightarrow \widehat{\mathbf{E}} = -\nabla \hat{\varphi} - \frac{\partial \widehat{\mathbf{A}}}{\partial t}, \qquad (18)$$

*as shown in Figure 7.*

Because uncountably many non-unique choices of A and φ generate the same **E** and **B** fields, we add the Lorenz gauge as a PINN-type soft constraint to the training cost function

$$w_B \|B - \hat{B}\|_2 + w_E \|E - \hat{E}\|_2 + w_L \left\|\nabla \cdot \widehat{A} + \frac{1}{c^2} \frac{\partial \hat{\varphi}}{\partial t}\right\|_2, (19)$$

which has the additional benefit that it introduces more data for the magnetic field calculation as the magnetic field is now informed by the Lorenz condition. Predictions for the entire 3D beam at step 1 by the Lorenz PCNN are shown in Figure 8. In Figure 9 we show the Lorenz PCNN-generated ($\mathbf{B}^\wedge$, $\mathbf{E}^\wedge$) fields at just a single 2D (x, y) slice at various steps including those beyond the training data.

## DISCUSSION

Our final demonstration of the strength of building in hard physics constraints in the 3D CNN is a demonstration of its non-catastrophic failure when predicting the electromagnetic fields of two additional 2 nC beams that are very different from both the test and training data shown so far. The first additional beam is three parallel electron beams, each of whose length is $\sigma_z \approx 3mm$ which is similar to the overall length of the various bunches used in the training data. The three parallel beams are different from the training data in having empty space between the individual bunches. The second additional beam is a hollow tube of electrons with the same length as the three parallel beams, but whose topology is entirely different from anything the PCNN has seen so far. Figure 10 shows results of predicting the (E, B) fields for the three parallel and the hollow beams. As expected, the PCNN performs worse than previously, but is qualitatively very accurate in both E and B field prediction for the three parallel beams. The hollow beam is a much bigger challenge and much larger field errors are seen, but crucially, the predicted fields are still qualitatively correct in terms of direction and flow, with most of the error due to the wrong amplitude being predicted. We believe that this final result honestly shows the generality and strength of the PCNN approach and its limitations. We should not expect a trained CNN to predict well on an entirely unseen data set, this is a well-known problem known as distribution shift in the ML community in which NNs must be re-trained for inputs different from the training data set distribution. The fact that the PCNN produces reasonable outputs for inputs wildly different from that of the training data is a major strength of the approach. As Maxwell's equations are important for describing an extremely wide range of physical phenomenon the applications of such a method for electrodynamics applications are many. Here we will briefly touch on charged particle dynamics in high energy accelerators. There is a growing literature on utilizing ML-based surrogate models as virtual diagnostics in the particle accelerator community. For these approaches the NNs are typically trained as input-output maps utilizing experimental input data together with computationally expensive output data such as measuring a charged particle current density at one location of an accelerator and then running physics codes to map that to another location [41], or mapping input accelerator data directly to beam characteristics at other accelerator locations [39]. For such applications, the PCNN method can enable the development of much more robust real-time virtual diagnostics that satisfy physics constraints.

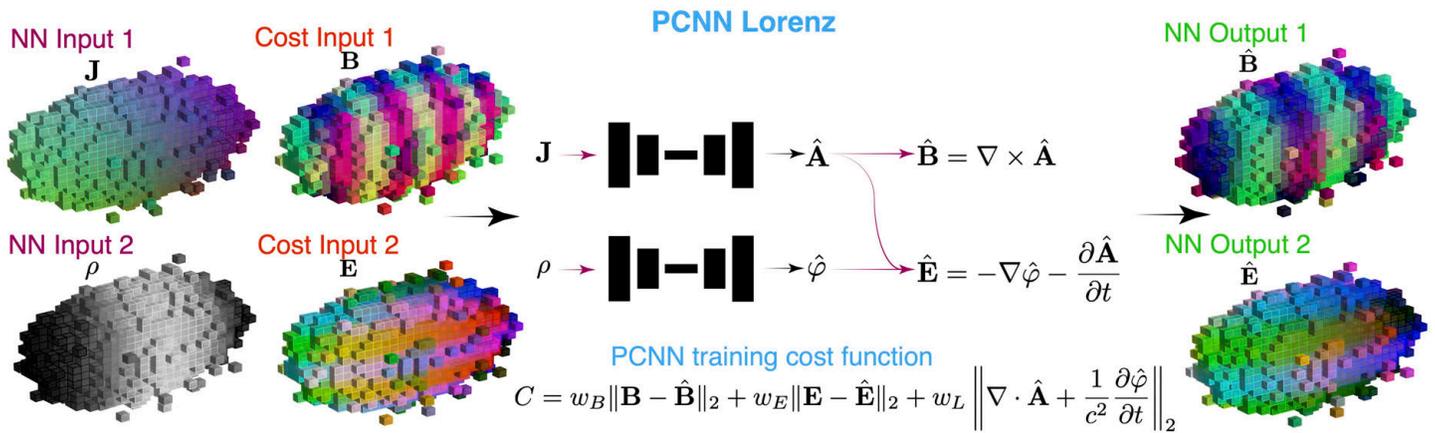

*Figure 7: The Lorenz PCNN generates the vector and scalar potentials and their associated electromagnetic fields.*

Another large family of applications is for accelerator design. For example, given a fixed input beam charge and current density distributions a beam line may be designed with various electromagnetic magnet and resonant acceleration components. For each design choice, such as distance between magnets or magnetic field strengths, high-fidelity physics-based models must be used to track the charged particle dynamics. With our approach, once a PCNN is trained for a family of input beam distributions, we have demonstrated that we can make accurate field predictions that respect physics constraints even as the beam is significantly changed by the application of external fields based on the accelerator's design. The next step of this work, which is beyond the scope of this paper and an ongoing effort, is to utilize our PCNN approach to quickly push particles and to confirm that the field predictions are accurate enough such that the particle dynamics are physically consistent. As we have already seen some slight numerical limitations as discussed above, this might push us to utilize even higher resolution discretization, such as $512^3$ or $1024^3$ pixel volumes, which remains to be determined. If this approach is able to provide physically consistent beam dynamics, even if they slightly violate constraints, this will be a fast and powerful way to zoom in on an optimal design estimate, after which more accurate slower physics-based simulations can be used for detailed studies.

**CONCLUSIONS**

A robust PCNN method has been developed in order to explicitly take Maxwell's equations into account in the structure of generative 3D convolutional neural networks in order to more accurately satisfy physics constraints. Although this method is less general than the incredibly flexible PINN approach, in which any partial differential equation can be easily introduced as a soft constraint, the resulting physics constraints are more accurately respected. Furthermore, we have shown how to combine this PCNN approach with the PINN approach in our Lorenz CNN in which hard physics con- straints are enforced in the generation of the E and B fields and the soft penalty on violation of the Lorenz guage is added to the cost function in Equation 19.

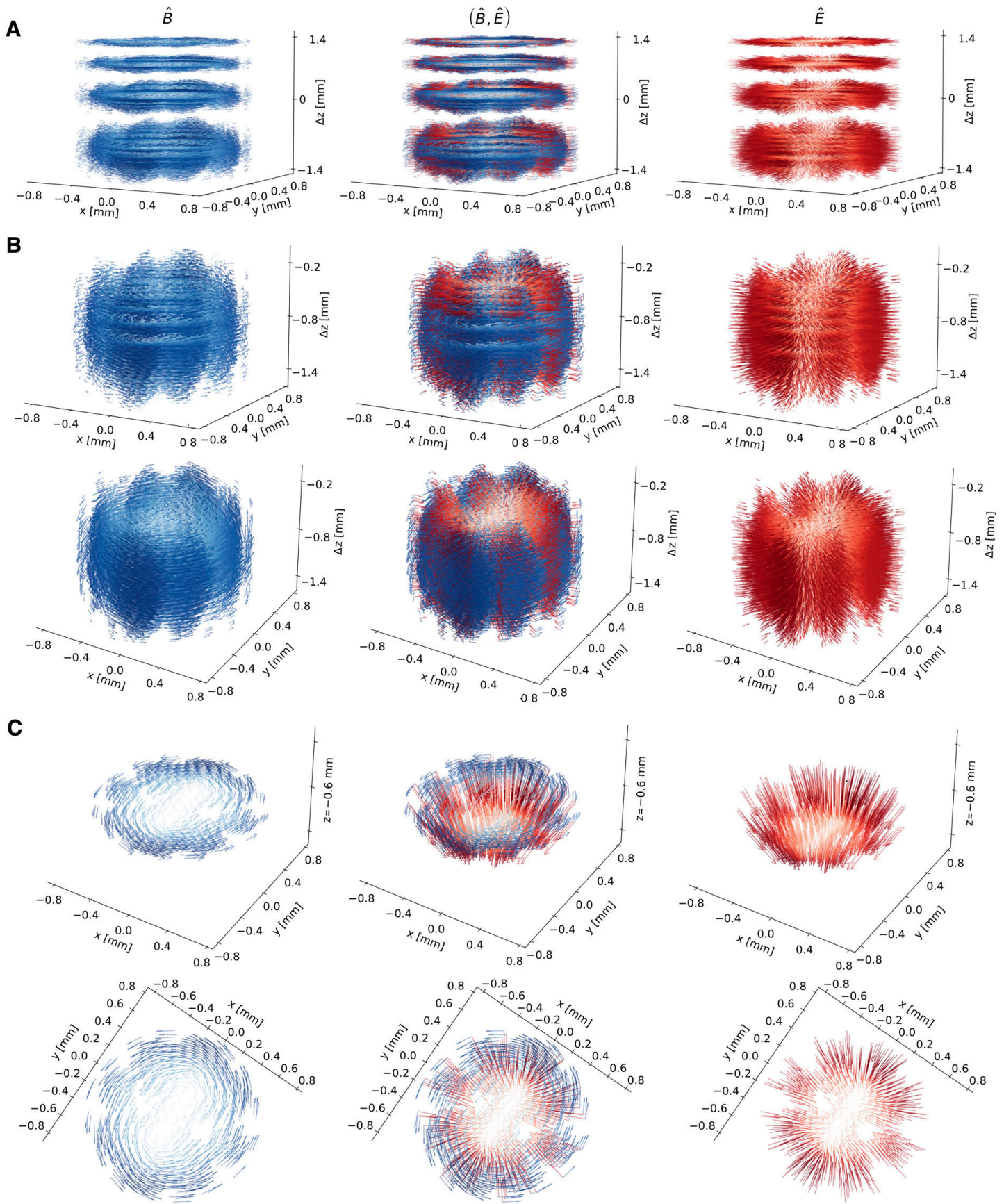

Fig 8: (A) Electromagnetic fields are shown for all positions within the $128^3$ pixel volume for normalized charge density $\rho >$ 0.0025 for the first state of the beam. (B) We zoom in on only the part of the electron bunch which has the largest $\sigma_z$ profile and show it from two angles. (C) Fields from only a single $(x, y)$ slice of the 3D volume are shown at two different angles.

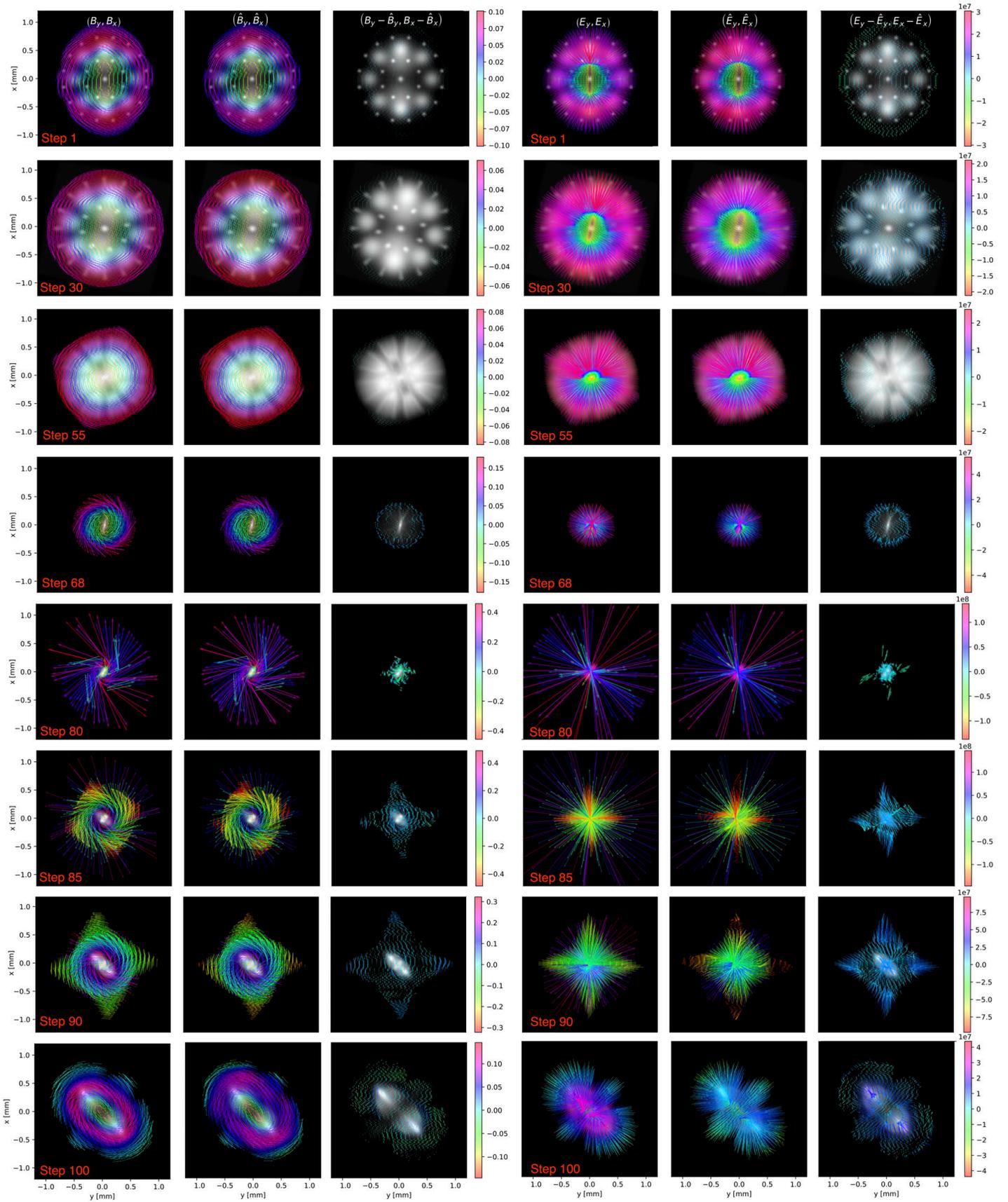

Fig 9: Lorenz PCNN generated electromagnetic fields are shown as the beam is over-compressed and then diverges again with the fields plotted over the (x,y) projection of the beam's charge density ρ.

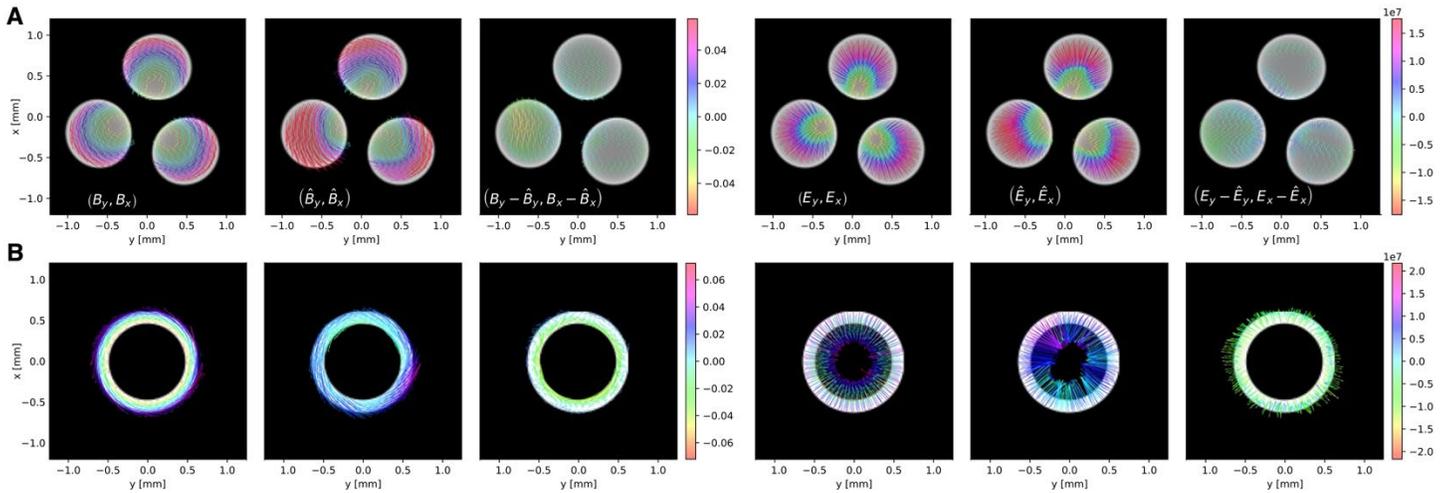

Figure 10: Projections into the (*x*,*y*) plane of the Lorenz PCNN-generated electromagnetic fields are shown for a pair of beams which are very different from any of the training data. (A) For a set of three parallel beams the field calculations are in good agreement. (B) For a hollow cylindrical beam, a completely different topology than anything seen by the PCNN in training, the amplitudes of the predicted fields show large errors, but qualitatively the field orientations are physical and match the correct fields.

## ACKNOWLEDGMENTS

This work was supported by the U.S. Department of En- ergy (DOE), Office of Science, Office of High Energy Physics contract number 89233218CNA000001 and the Los Alamos National Laboratory LDRD Program Directed Research (DR) project 20220074DR. National Laboratory LDRD Program Directed Research (DR) project 20220074DR.

## REFERENCES

[1] V. De Lorenci, R. Klippert, M. Novello, and J. Salim, "Nonlinear electro- dynamics and frw cosmology," Physical Review D 65, 063501 (2002).

[2] M. Novello, S. P. Bergliaffa, and J. Salim, "Nonlinear electrodynamics and the acceleration of the universe," Physical Review D 69, 127301 (2004).

[3] S. Kruglov, "Universe acceleration and nonlinear electrodynamics," Physical Review D 92, 123523 (2015).

[4] I. Bandos, K. Lechner, D. Sorokin, and P. K. Townsend, "Nonlinear duality-invariant conformal extension of maxwell's equations," Physical Review D 102, 121703 (2020).

[5] J. Li, S. Kamin, G. Zheng, F. Neubrech, S. Zhang, and N. Liu, "Addressable metasurfaces for dynamic holography and optical information encryption," Science advances 4, eaar6768 (2018).

[6] C. Evoli, P. Blasi, G. Morlino, and R. Aloisio, "Origin of the cosmic ray galactic halo driven by advected turbulence and self-generated waves," Physical Review Letters 121, 021102 (2018).

[7] J. Nättilä and A. M. Beloborodov, "Heating of magnetically dominated plasma by alfvén-wave turbulence," Physical Review Letters 128, 075101 (2022).

[8] X. Luo, Y. Hu, X. Ou, X. Li, J. Lai, N. Liu, X. Cheng, A. Pan, and H. Duan, "Metasurface-enabled on-chip multiplexed diffractive neural networks in the visible," Light: Science & Applications 11, 1–11 (2022).

[9] F. Aieta, M. A. Kats, P. Genevet, and F. Capasso, "Multiwavelength achro- matic metasurfaces by dispersive phase compensation," Science 347, 1342– 1345 (2015).

[10] J. Miao, T. Ishikawa, I. K. Robinson, and M. M. Murnane, "Beyond crystal- lography: Diffractive imaging using coherent x-ray light sources," Science 348, 530–535 (2015).

[11] F. Bacchini, F. Pucci, F. Malara, and G. Lapenta, "Kinetic heating by alfvén waves in magnetic shears," Physical Review Letters 128, 025101 (2022).

[12] C. Chaston, C. Carlson, W. Peria, R. Ergun, and J. McFadden, "Fast observations of inertial alfvén waves in the dayside aurora," Geophysical research letters 26, 647–650 (1999).

[13] J. Deng, Y. H. Lo, M. Gallagher-Jones, S. Chen, A. Pryor Jr, Q. Jin, Y. P. Hong, Y. S. Nashed, S. Vogt, J. Miao, *et al.*, "Correlative 3d x-ray fluorescence and ptychographic tomography of frozen-hydrated green algae," Science advances 4, eaau4548 (2018).

[14] H. Gurram, J. Egedal, and W. Daughton, "Shear Alfvén waves driven by magnetic reconnection as an energy source for the aurora borealis," Geo- physical Research Letters 48, e2021GL094201 (2021).

[15] S. Tang, T. Carter, N. Crocker, W. Heidbrink, J. Lestz, R. Pinsker, K. Thome, M. Van Zeeland, and E. Belova, "Stabilization of Alfvén eigen- modes in diii-d via controlled energetic ion density ramp and validation of theory and simulations," Physical Review Letters 126, 155001 (2021).

[16] H. Tang, L. Zhao, P. Zhu, X. Zou, J. Qi, Y. Cheng, J. Qiu, X. Hu, W. Song, D. Xiang, *et al.*, "Stable and scalable multistage terahertz-driven particle accelerator," Physical Review Letters 127, 074801 (2021).

[17] Z. Zhang and S. Satpathy, "Electromagnetic wave propagation in periodic structures: Bloch wave solution of maxwell's equations," Physical review letters 65, 2650 (1990).

[18] W. Zachariasen, "A general theory of x-ray diffraction in crystals," Acta Crystallographica 23, 558–564 (1967).

[19] L. Monico, L. Cartechini, F. Rosi, A. Chieli, C. Grazia, S. De Meyer, G. Nuyts, F. Vanmeert, K. Janssens, M. Cotte, *et al.*, "Probing the chemistry of cds paints in the scream by in situ noninvasive spectroscopies and synchrotron radiation x-ray techniques," Science advances 6, eaay3514 (2020).

[20] P. Mehrabi, R. Bücker, G. Bourenkov, H. Ginn, D. von Stetten, H. Müller- Werkmeister, A. Kuo, T. Morizumi, B. Eger, W.-L. Ou, *et al.*, "Serial fem tosecond and serial synchrotron crystallography can yield data of equivalent quality: A systematic comparison," Science advances 7, eabf1380 (2021).

[21] F. Hartemann and N. Luhmann Jr, "Classical electrodynamical derivation of the radiation damping force," Physical review letters 74, 1107 (1995). [22]D. Grošelj, A. Mallet, N. F. Loureiro, and F. Jenko,

"Fully kinetic simulation of 3d kinetic Alfvén turbulence," Physical review letters 120, 105101 (2018).

[23] G. Franchetti, I. Hofmann, and D. Jeon, "Anisotropic free-energy limit of halos in high-intensity accelerators," Physical review letters 88, 254802 (2002).

[24] G. Franchetti, I. Hofmann, and M. Aslaninejad, "Collective emittance exchange with linear space charge forces and linear coupling," Physical review letters 94, 194801 (2005).

[25] Y. Cai, "Coherent synchrotron radiation by electrons moving on circular orbits," Physical Review Accelerators and Beams 20, 064402 (2017).

[26] I. Zagorodnov, S. Tomin, Y. Chen, and F. Brinker, "Experimental valida-tion of collective effects modeling at injector section of x-ray free-electron laser," Nuclear Instruments and Methods in Physics Research Section A: Accelerators, Spectrometers, Detectors and Associated Equipment 995,
165111 (2021).

[27] P. Emma, R. Akre, J. Arthur, R. Bionta, C. Bostedt, J. Bozek, A. Brachmann, P. Bucksbaum, R. Coffee, F.-J. Decker, *et al.*, "First lasing and oper- ation of an rangstrom-wavelength free-electron laser," nature photonics 4, 641 (2010).

[28] C. Milne, T. Schietinger, M. Aiba, A. Alarcon, J. Alex, A. Anghel, V. Arsov, C. Beard, P. Beaud, S. Bettoni, *et al.*, "Swissfel: the Swiss x-ray free electron laser," Applied Sciences 7, 720 (2017).

[29] W. Decking, S. Abeghyan, P. Abramian, A. Abramsky, A. Aguirre, C. Al- brecht, P. Alou, M. Altarelli, P. Altmann, K. Amyan, *et al.*, "A MHZ- repetition-rate hard x-ray free-electron laser driven by a superconducting linear accelerator," Nature Photonics 14, 391–397 (2020).

[30] S. Gessner, E. Adli, J. M. Allen, W. An, C. I. Clarke, C. E. Clayton, S. Corde, J. Delahaye, J. Frederico, S. Z. Green, *et al.*, "Demonstration of a positron beam-driven hollow channel plasma wakefield accelerator," Nature communications 7, 1–6 (2016).

[31] V. Yakimenko, L. Alsberg, E. Bong, G. Bouchard, C. Clarke, C. Emma, S. Green, C. Hast, M. Hogan, J. Seabury, *et al.*, "FACET-II facility for ad vanced accelerator experimental tests," Physical Review Accelerators and Beams 22, 101301 (2019).

[32] Y. Chen, I. Zagorodnov, and M. Dohlus, "Beam dynamics of realistic bunches at the injector section of the European x-ray free-electron laser," Physical Review Accelerators and Beams 23, 044201 (2020).

[33] S. Van Der Geer, O. Luiten, M. De Loos, G. Pöplau, and U. Van Rienen, "3d space-charge model for GPT simulations of high brightness electron bunches," in *Institute of Physics Conference Series*, Vol. 175 (2005) p. 101.

[34] A. Brynes, P. Smorenburg, I. Akkermans, E. Allaria, L. Badano, S. Brus- saard, M. Danailov, A. Demidovich, G. De Ninno, D. Gauthier, *et al.*, "Beyond the limits of 1d coherent synchrotron radiation," New Journal of Physics 20, 073035 (2018).

[35] M. Dohlus, T. Limberg, *et al.*, "Csrtrack: faster calculation of 3d CSR effects," in *Proceedings of the 2004 FEL Conference* (2004) pp. 18–21.

[36] S. Li, P. M. Dee, E. Khatami, and S. Johnston, "Accelerating lattice quantum Monte Carlo simulations using artificial neural networks: Application to the Holstein model," Physical Review B 100, 020302 (2019).

[37] A. Scheinker and R. Pokharel, "Adaptive 3d convolutional neural network- based reconstruction method for 3d coherent diffraction imaging," Journal of Applied Physics 128, 184901 (2020).

[38] E. Rrapaj and A. Roggero, "Exact representations of many-body interactions with restricted-Boltzmann-machine neural networks," Physical Re- view E 103, 013302 (2021).

[39] J. Zhu, Y. Chen, F. Brinker, W. Decking, S. Tomin, and H. Schlarb, "High- fidelity prediction of megapixel longitudinal phase-space images of electron beams using encoder-decoder neural networks," Physical Review Applied 16, 024005 (2021).

[40] J. Zhu, N. Lockmann, M. Czwalinna, and H. Schlarb, "Mixed diagnostics for longitudinal properties of electron bunches in a free-electron laser," arXiv preprint arXiv:2201.05769 (2022), https://doi.org/10.48550/arXiv.2201.05769.

[41] A. Scheinker, "Adaptive machine learning for time-varying systems: low dimensional latent space tuning," Journal of Instrumentation 16, P10008
(2021).

[42] A. Scheinker, F. Cropp, S. Paiagua, and D. Filippetto, "An adaptive approach to machine learning for compact particle accelerators," Scientific reports 11, 1–11 (2021).

[43] E. Fol, R. Tomás, and G. Franchetti, "Supervised learning-based reconstruction of magnet errors in circular accelerators," The European Physical Journal Plus 136, 365 (2021).

[44] M. Raissi, A. Yazdani, and G. E. Karniadakis, "Hidden fluid mechanics: Learning velocity and pressure fields from flow visualizations," Science 367, 1026–1030 (2020).

[45] E. Zhang, M. Dao, G. E. Karniadakis, and S. Suresh, "Analyses of internal structures and defects in materials using physics-informed neural networks," Science advances 8, eabk0644 (2022).

[46] A. Ivanov and I. Agapov, "Physics-based deep neural networks for beam dynamics in charged particle accelerators," Physical review accelerators and beams 23, 074601 (2020).

[47] S. Chmiela, A. Tkatchenko, H. E. Sauceda, I. Poltavsky, K. T. Schütt, and K.-R. Müller, "Machine learning of accurate energy-conserving molecular force fields," Science advances 3, e1603015 (2017).

[48] G. E. Karniadakis, I. G. Kevrekidis, L. Lu, P. Perdikaris, S. Wang, and L. Yang, "Physics-informed machine learning," Nature Reviews Physics 3, 422–440 (2021).

[49] L. Lu, P. Jin, G. Pang, Z. Zhang, and G. E. Karniadakis, "Learning nonlinear operators via deeponet based on the universal approximation theorem of operators," Nature Machine Intelligence 3, 218–229 (2021).

[50] S. Wang, H. Wang, and P. Perdikaris, "Learning the solution operator of parametric partial differential equations with physics-informed deeponets," Science advances 7, eabi8605 (2021).

[51] T. J. Bridges and S. Reich, "Multi-symplectic integrators: numerical schemes for Hamiltonian PDEs that conserve symplecticity," Physics Letters A 284, 184–193 (2001).

[52] I. Zagorodnov, R. Schuhmann, and T. Weiland, "Long-time numerical computation of electromagnetic fields in the vicinity of a relativistic source," Journal of Computational Physics 191, 525–541 (2003).

[53] J. Frank, B. E. Moore, and S. Reich, "Linear PDEs and numerical methods that preserve a multisymplectic conservation law," SIAM Journal on Scientific Computing 28, 260–277 (2006).

[54] Y. Cai, "Symplectic maps and chromatic optics in particle accelerators," Nuclear Instruments and Methods in Physics Research Section A: Accelerators, Spectrometers, Detectors and Associated Equipment 797, 172–181 (2015).

[55] J. Qiang, "Symplectic multiparticle tracking model for self-consistent space-charge simulation," Physical Review Accelerators and Beams 20, 014203 (2017).

[56] J. Qiang, "Symplectic particle-in-cell model for space-charge beam dynamics simulation," Physical Review Accelerators and Beams 21, 054201 (2018).

[57] J. Ling, A. Kurzawski, and J. Templeton, "Reynolds averaged turbulence modelling using deep neural networks with embedded invariance," Journal of Fluid Mechanics 807, 155–166 (2016).

[58] J.-L. Wu, H. Xiao, and E. Paterson, "Physics-informed machine learning approach for augmenting turbulence models: A

comprehensive frame- work," Physical Review Fluids 3, 074602 (2018).

[59] A. T. Mohan, N. Lubbers, D. Livescu, and M. Chertkov, "Embedding hard physical constraints in convolutional neural networks for 3d turbulence," in *ICLR 2020 Workshop on Integration of Deep Neural Models and Differential Equations* (2019).

[60] L. Sun, H. Gao, S. Pan, and J.-X. Wang, "Surrogate modeling for fluid flows based on physics-constrained deep learning without simulation data," Computer Methods in Applied Mechanics and Engineering 361, 112732 (2020).

[61] T. Beucler, M. Pritchard, S. Rasp, J. Ott, P. Baldi, and P. Gentine, "Enforcing analytic constraints in neural networks emulating physical systems," Physical Review Letters 126, 098302 (2021).

[62] M. Abadi, A. Agarwal, P. Barham, E. Brevdo, Z. Chen, C. Citro, G. S. Corrado, A. Davis, J. Dean, M. Devin, *et al.*, "Tensorflow: Large-scale machine learning on heterogeneous systems," (2015).

[63] A. M. de la Ossa, T. Mehrling, and J. Osterhoff, "Intrinsic stabilization of the drive beam in plasma-wakefield accelerators," Physical review letters 121, 064803 (2018).